\documentclass[a4paper,12pt]{article}
\usepackage{amsmath}
\usepackage{amsfonts}
\usepackage{amssymb}
\usepackage{graphicx}
\usepackage{cite}

\begin{document}

\date{}

\title{\bf Anisotropic dark energy model with a hybrid scale factor}

\author{B. Mishra \footnote{Department of Mathematics, Birla Institute of Technology and Science-Pilani,Hyderabad Campus,Hyderabad-500078, INDIA, bivudutta@yahoo.com} 
and 
S. K. Tripathy\footnote{Department of Physics, Indira Gandhi Institute of Technology, Sarang, Dhenkanal, Odisha-759146, INDIA, tripathy\_ sunil@rediffmail.com}      
}

\maketitle

\begin{abstract} Anisotropic dark energy model with dynamic pressure anisotropies along different spatial directions  is constructed at the backdrop of a spatially homogeneous diagonal Bianchi type $V$ $(BV)$ space-time in the framework of General Relativity. A time varying deceleration parameter generating a hybrid scale factor is considered to simulate a cosmic transition from early deceleration to late time acceleration. We found that the pressure anisotropies  along the $y-$ and $z-$ axes evolve dynamically and continue along with the cosmic expansion without being subsided even at late times. The anisotropic pressure along the $x-$axis becomes equal to the mean fluid pressure. At a late phase of cosmic evolution, the model enters into a phantom region. From a statefinder diagnosis, it is found that the model overlaps with $\Lambda$CDM at late phase of cosmic  time.
\end{abstract}



\section{Introduction}

It is now an accepted fact that, the universe is undergoing an accelerated phase of expansion in the present epoch. A lot of observational data support this fact \cite{Riess98, Perl99, Spergel07, Komatsu09, Tegmark04, Seljak05, Eisen05}. The accelerated expansion is believed to be a late time dynamics of the universe. Strong observational evidences have led to a wide consensus that the transition from a decelerated phase to an accelerated one occurred at a transition redshift $z_{da}\sim 1$ \cite{Farooq13, Busca13, Capo14}. The reason behind this late time dynamics is not exactly known. However this phenomena is attributed to an exotic dark energy(DE) form which must have a lion share of $68.3\%$ in the mass-energy budget to account for the acceleration \cite{Ade14, Ade14a, Ade14b}. In General Relativity, dark energy corresponds to an isotropic fluid with almost constant energy density with negative pressure. The late time dynamics of the universe triggered enormous research works with novel concepts and ideas. Dark energy can be better understood through an equation of state parameter $\omega$ defined as the ratio $\omega = \frac{p}{\rho}$, where $p$ is the pressure and $\rho$ is energy density. Under the purview of General Relativity, there have been a good number of models proposed to understand the nature and behaviour of DE. Besides the consideration of a cosmological constant with $\omega = -1$ ( $\Lambda$CDM model), canonical scalar field models such as quintessence ($-\frac{2}{3}\leq \omega \leq -\frac{1}{3} $) \cite{Ratra88,Sahni00}, phantom fields ($\omega < -1$) \cite{Cald02}, k-essence \cite{Picon00, Picon01}, tachyons \cite{Sen02}, quintom \cite{Feng05, Guo05} have been suggested. Alternative dark energy candidates such as ghost dark energy \cite{Urban09, Urban10, Ohta11}, holographic dark energy \cite{Li04},  Ricci dark energy \cite{Cao09} and agegraphic dark energy \cite{Cai07, Wei08} have been proposed in recent times where a parametrized form of the dark energy density is considered. Basing upon the requirements, the parametrized forms are tuned to get viable models describing dark energy. In recent times, there have also been a growing interest in a unified dark fluid where the contributions coming from both the dark energy and dark matter part are handled through a unified dark equation of state which may be linear or non linear in energy density \cite{Anand06, Balbi07, Xu12, Liao12, SKT15a}. The single unified dark adiabatic cosmic fluid is able to explain most of the recent observational data. Another approach is the  modification of the Einstein-Hilbert action of Einstein theory which has also been proved to be successful in providing some insights into the dark energy problem \cite{Caroll04, Nojiri06, Nojiri11, Harko11, Bamba12, Bamba15}. Even though, the dark energy equation of state is considered to be a constant quantity in most of these models, it is not very much necessary and should be allowed to vary with time. Of course in certain models, this parameter comes out to be evolving with cosmic dynamics.  

The universe is mostly observed to be flat and isotropic and supports the predictions of $\Lambda$CDM model. However, observations of high resolution CMB radiation data from Wilkinson Microwave Anisotropy Probe (WMAP) showing some large angle anomalies  suggest a non-trivial topology of the large scale geometry of the universe with an asymmetric expansion \cite{Hinshaw07, Hinshaw09, Costa04, Watan09}.  Planck data also show a slight redshift of the primordial power spectrum of curvature perturbation from exact scale invariance \cite{Ade14}. These observations obviously hint towards the presence of some anisotropic energy source in the universe with anisotropic pressures. The issue of global anisotropy can be settled if anisotropy can be incorporated to the flat Friedman model (FRW) as a sort of small perturbation or if the FRW models can be modified in a suitable manner. In order to address the issue of the smallness in the angular power spectrum, some anisotropic models have been proposed in recent times \cite{Campa06, Campanel07, Campa09, Gruppo07}. These models bear a similarity to the Bianchi morphology \cite{Jaffe05, Jaffe06, Jaffe06a}. Spatially homogeneous Bianchi type models are more general than the FRW models and  have anisotropic spatial sections. They provide an opportunity to consider asymmetric expansion along different spatial sections. 

In some recent works \cite{BM15, SKT15}, we have investigated the background cosmologies of some dark energy models with anisotropic pressures in the backdrop of anisotropic $BV$ metric in the framework of  a scale invariant theory as proposed by Wesson \cite{Wesson81a, Wesson81b}. Keeping in view of the late time dynamics of the universe, we considered a constant deceleration parameter which provides two volumetric expansion behaviour of the universe namely power law and exponential expansion. It has been shown in those works that, there remain pressure anisotropies even at late phase of cosmic evolution. Also, we have investigated the role played by the scale invariance in the coordinates in comparison to that in the absence of scale invariance. It is worth to note that, the scale invariant formulation we have adotpted reduce to General Relativity without cosmological constant under the assumption of a time independent Dirac gauge function.  In the present work, we intend to extend those works by incorporating the time varying behaviour of deceleration parameter so that it can mimic a cosmic transition. 

The paper is organised as follows: In Section 2, the basic formalism for anisotropic dark energy model with anisotropic pressures along different spatial directions have been discussed for an anisotropic and spatially homogeneous $BV$ metric in the framework of General Relativity. Similar formalism has already been developed in our earlier works \cite{BM15, SKT15}. Concept of pressure anisotropy is not new and has been investigated in literature. In a recent work, Akarsu has considered such pressure anisotropy for a $BIII$ metric to investigate the isotropisation of the model at late times \cite{Akarsu10}. In Section 3, a hybrid scale factor having both exponential and power law nature generated by a time varying deceleration parameter is considered to mimic a cosmic transition. The deceleration parameter has a characteristic to decrease from a positive value at early time to an asymptotic negative value at late time of cosmic evolution.  The formulations for the skewness parameters and pressure anisotropies are derived. The dynamics of the model  and the dynamics of the pressure anisotropies  in the form of the skewness parameters are discussed. The viability of the discussed dark energy model is tested with a statefinder diagnosis in Section 4. At the end, the conclusions of the work are presented in Section 5.

\section{Basic Formalism}
We consider an anisotropic dark energy model with anisotropic pressures along different spatial directions in the field equations in General Relativity, $G_{ij} \equiv R_{ij}-\frac{1}{2}Rg_{ij}=-T_{ij}$, where the energy momentum tensor for dark energy is assumed as 
\begin{eqnarray}\label{eq:1}
T_{ij} &=& diag[\rho, -p_x, -p_y,-p_z]\nonumber \\
          &=& diag[1, -\omega_x, -\omega_y,-\omega_z]\rho\nonumber\\
          &=&diag[1,-(\omega +\delta), -(\omega+\gamma), -(\omega+\eta)]\rho.
\end{eqnarray}
The skewness parameters $\delta$, $\gamma$ and $\eta$ are the respective deviations along $x-$, $y-$ and $z$ axes from the equation of state (EoS) parameter $\omega$. We allow these skewness parameters to evolve with the cosmic dynamics. $\rho$ is the energy density and the pressure $p=\omega \rho$. Here we have used the gravitational units $(8\pi G=c=1)$.  The line element for Bianchi type $V$ (BV) space-time is considered in the form  
\begin{equation}\label{eq:2}
ds^{2}=-dt^{2}+A^{2}dx^{2}+e^{2 \alpha x}(B^{2}dy^{2}+C^{2}dz^{2}).
\end{equation}
where the directional scale factors $A=A(t), B=B(t), C=C(t)$ are functions of cosmic time only and $\alpha$ is a positive constant. Einstein field equations for the metric \eqref{eq:2} are 
\begin{equation}\label{eq:3}
\frac{\ddot{B}}{B}+\frac{\ddot{C}}{C}+\frac{\dot{B}\dot{C}}{BC}-\frac{\alpha^2}{A^2}= -(\omega+\delta)\rho,
\end{equation}

\begin{equation}\label{eq:4}
\frac{\ddot{A}}{A}+\frac{\ddot{C}}{C}+\frac{\dot{A}\dot{C}}{AC}-\frac{\alpha^2}{A^2}= -(\omega+\gamma)\rho,
\end{equation}

\begin{equation}\label{eq:5}
\frac{\ddot{A}}{A}+\frac{\ddot{B}}{B}+\frac{\dot{A}\dot{B}}{AB}-\frac{\alpha^2}{A^2}= -(\omega+\eta)\rho,
\end{equation}

\begin{equation}\label{eq:6}
\frac{\dot{A}\dot{B}}{AB}+\frac{\dot{B}\dot{C}}{BC}+\frac{\dot{A}\dot{C}}{AC}-3\frac{\alpha^2}{A^2}=\rho
\end{equation}

and

\begin{equation}\label{eq:7}
2\frac{\dot{A}}{A}-\frac{\dot{B}}{B}-\frac{\dot{C}}{C}=0.
\end{equation}

An overhead dot on a field variable denotes differentiation with respect to time $t$. On integration, eqn. \eqref{eq:7} yields   
\begin{equation}\label{eq:8}
A^2= BC,
\end{equation}
where the integration constant is taken to be 1.

The energy conservation for the anisotropic fluid, $T^{ij}_{;j}=0 $, yields

\begin{equation}\label{eq:9}
\dot{\rho}+3\rho(\omega+1)H+\rho(\delta H_x+\gamma H_y+\eta H_z)=0,
\end{equation}
where the directional Hubble rates are defined as $H_x=\frac{\dot{A}}{A}$, $H_y=\frac{\dot{B}}{B}$ and $H_z=\frac{\dot{C}}{C}$ and the mean Hubble rate is $H=\frac{1}{3}\left(H_x+H_y+H_z\right)$. The above equation \eqref{eq:9} can be split into two parts: the first one corresponds to the conservation of matter field with equal pressure along all the directions i.e. the deviation free part of \eqref{eq:9}  and the second one corresponds to that involving the deviations of EoS parameter:
\begin{equation}\label{eq:10}
\dot{\rho}+3\rho(\omega+1)H=0,
\end{equation}
and
\begin{equation}\label{eq:11}
\rho(\delta H_x+\gamma H_y+\eta H_z)=0.
\end{equation}
It is now certain that, the behaviour of the energy density $\rho$ is controlled by the deviation free part of EoS parameter whereas  the anisotropic pressures along different spatial directions can be obtained from the second part of the conservation equation. From equation \eqref{eq:10}, we obtain the energy density as $\rho=\rho_0 \mathcal{R}^{-3(\omega+1)}$, where $\rho_0$ is the value of energy density at the present epoch and $\mathcal{R}$ is the scale factor of the universe.

The Friedman equivalent equation for the field equations \eqref{eq:3}-\eqref{eq:6} can be expressed as 

\begin{equation}\label{eq:12}
p+\frac{1}{3}(3\omega+\delta+\gamma+\eta)\rho=H^2(2q-1)-\sigma^2+\frac{\alpha^2}{A^2}.
\end{equation}
where we have used the expression $q=-1-\frac{\dot{H}}{H^2}$ for deceleration parameter. The scalar expansion $\theta$ and shear scalar $\sigma^2$ in the model are expressed as
\begin{equation}\label{eq:13}
\theta=3H=\frac{\dot{A}}{A}+\frac{\dot{B}}{B}+\frac{\dot{C}}{C},
\end{equation}

and

\begin{equation}\label{eq:14}
\sigma^2=\frac{1}{2}\sigma_{ij}\sigma^{ij}=\frac{1}{2}\biggl(\Sigma H_i^2-\frac{1}{3}\theta^2\biggr),
\end{equation}
where $H_i; i=1,2,3$ are the respective directional Hubble rates along $x-$, $y-$ and $z-$ axes. Also, $\sigma_{ij}=\frac{1}{2}(u_{i;k}h_j^k+u_{j;k}h_i^k-\frac{1}{3}\theta h_{ij})$ and $ h_{ij}=g_{ij}-u_iu_j $ is the projection tensor. $u_i= \delta_i^4$ is the four velocity vector in the comoving coordinates. The shear scalar is usually considered to be proportional to the scalar expansion for spatially homogeneous metrics which leads to an anisotropic relationship among the directional scale factors $B$ and $C$ as $B=C^m$ \cite{Collins80, SKT10, BM15}. Here the exponent $m$ is a positive constant and takes care of the anisotropic nature of the model. The model is isotropic for $m=1$ else anisotropic.

In the present work, we assume the rate of expansion along the $x-$axis to be the same as that of the mean expansion rate i.e $H_x=H$. This will lead to the consideration that, the pressure along $x-$axis is mostly the same as that of the total pressure. The directional Hubble rates along other two spatial directions can now be expressed as

\begin{eqnarray}\label{eq:17}
H_y &=& \left(\frac{2m}{m+1}\right)H ,\nonumber\\ 
H_z &=& \left(\frac{2}{m+1}\right)H.
\end{eqnarray}

Following the formalism developed in Refs. \cite{BM15,SKT15}, the skewness parameters can be obtained as

\begin{eqnarray}
\delta &=& -\left(\frac{m-1}{3\rho}\right)\zeta(m) F(H),\label{eq:18}\\
\gamma &=& \left(\frac{5+m}{6\rho}\right)\zeta (m) F(H),\label{eq:19}\\
\eta &=& -\left(\frac{5m+1}{6\rho}\right)\zeta (m) F(H),\label{eq:20}
\end{eqnarray}
where, $\zeta (m)=\frac{m-1}{m+1}$ and $F(H)=\frac{2}{m+1}\left(\dot{H}+3H^2\right)$.  The functional $\zeta (m)$ measures the deviation from isotropic nature. For $m=1$, $\zeta (m)$ vanishes and consequently the cosmic fluid becomes isotropic. From equations \eqref{eq:18}-\eqref{eq:20}, it is clear that, the evolution of the pressure anisotropies or the skewness parameters are decided by the evolving nature of the factor $\frac{F(H)}{\rho}$. In General Relativity, the functional $F(H)$ has a great role in the description of the late time acceleration of the universe. One should note that, if the functional $F(H)$ vanishes then it will lead to the vanishing of the skewness parameters. Also, for a vanishing functional $F(H)$, the deceleration parameter becomes $q=2$, leading to a prediction of decelerating universe. It is worth to mention here that a positive deceleration parameter signifies a decelerating universe whereas its negative value indicates an accelerated cosmic expansion. In other words, a non vanishing $F(H)$ is required for the description of an accelerated expansion. In some earlier works \cite{SKT13, SKT14}, it has been shown that this functional in the form $\chi (H)= \dot{H}+3H^2$ vanishes for LRSBI models. Hence LRSBI models can not provide accelerating models without the contributions coming from magnetic field, scalar field or cosmic string. However, in presence of cosmic strings, magnetic field or scalar fields the situation gets modified and accelerating models are achieved \cite{SKT15a,SKT15b}. In the present work, we are interested in the late time cosmic dynamics with preconceived idea of accelerated expansion and therefore, we will force this functional to be non zero. In the background of this assumption, we will investigate the cosmic dynamics through the evolution of dark energy equation of state.  

The energy density $\rho$ and the EoS parameter $\omega$ are obtained as

\begin{eqnarray}
\rho &=& \frac{2(m^2+4m+1)H^2}{(m+1)^2}-\frac{3\alpha ^2}{\mathcal{R}^2},\label{eq:21}\\
\omega \rho &=& -\frac{2}{3}\left(\frac{m^2+4m+1}{m+1}\right)\left[F(H)-\frac{3H^2}{m+1}\right]+ \frac{\alpha ^2}{\mathcal{R}^2},\label{eq:22}
\end{eqnarray}
where the average scale factor is $\mathcal{R}=(ABC)^\frac{1}{3}$ which ultimately be the same as $A$. The above equations \eqref{eq:21} and \eqref{eq:22} clearly show that, if the expansion history is tracked by choosing a scale factor or more specifically a Hubble parameter, then the background cosmology can be easily studied. In our recent works \cite{BM15, SKT15}, we have followed similar approach to investigate the cosmic dynamics in a scale invariant theory of gravitation where we have considered a constant deceleration parameter simulating two different volumetric expansion laws namely power law expansion and exponential expansion. However, according to observations, the accelerated expansion of the universe is a recent phenomena which fosters the idea that, the universe might have undergone a transit at some point of time from a decelerated phase to an accelerated phase. A constant deceleration parameter can not predict this particular feature of cosmic expansion. In view of this, in the present work, we wish to consider a dynamically changing deceleration parameter which shows a behaviour of early deceleration with positive value and with the growth of cosmic time it switches over to a negative value predicting a late time acceleration.

\section{Cosmic transit and Hybrid scale factor}
A cosmic transit from early deceleration to late time acceleration can be obtained by a hybrid scale factor $\mathcal{R}=e^{at}t^b$ where $a$ and $b$ are positive constants. This scale factor has two factors: one factor behaving like exponential expansion and the other factor behaving like power law expansion. While the power law behaviour dominate the cosmic dynamics in early phase of cosmic evolution, the exponential factor dominates at late phase. When $b=0$, the exponential law is recovered and for $a=0$, the scale factor reduces to the power law.  The Hubble parameter for this model is $H=a+\frac{b}{t}$ and the directional Hubble parameters are $H_x=a+\frac{b}{t}$, $H_y=\frac{2m}{m+1}\left(a+\frac{b}{t}\right)$ and $H_z=\frac{2}{m+1}\left(a+\frac{b}{t}\right)$. Similar expansion law has already been conceived earlier \cite{Pradhan14, Akarsu14, Suresh13, SKT14}. In Ref. \cite{SKT14}, a more general form of such  hybrid Hubble parameter has been considered with the form $H=a+\frac{b}{t^n}$, $n$ being a constant. The present hybrid scale factor is a special case ($n=1$) of that considered in Ref. \cite {SKT14}. Consequently, the deceleration parameter becomes $q=-1+\frac{b}{(at+b)^2}$. At an early phase of cosmic evolution when $t\longrightarrow 0$, $q \simeq -1+\frac{1}{b}$ and at late phase of cosmic evolution with $t\longrightarrow \infty$, $q \simeq -1$. We are very much interested in a transient universe with early deceleration and late acceleration and therefore constrain the parameter $b$ to be in the range $0<b<1$ so that at early time $q$ can be positive whereas at late time $q$ assumes a negative value in conformity with the recent observational data. From the expression of the deceleration parameter we can infer that, the cosmic transit occurs at a time $t=-\frac{b}{a}\pm \frac{\sqrt{b}}{a}$. The negativity of the second term leads to a concept of negative time which may be unphysical in the context of Big Bang cosmology and therefore, the cosmic transit may have occurred at a time $t=\frac{\sqrt{b}-b}{a}$ which again restricts $b$ in the same range $0<b<1$.

The functional $F(H)$ for the hybrid scale factor becomes
\begin{equation}
F(t)=\frac{2}{m+1}\left[3a^2+\frac{6ab}{t}+\frac{(3b-1)b}{t^2}\right].\label{eq:23}
\end{equation}
We require that $F(t)$ should not vanish, at least for large cosmic time, so that we will get an accelerated expansion at late times of cosmic evolution. At late times, $F(t)\approx \frac{6a^2}{m+1}$ and is always having a non zero positive value since both $a$ and $m$ are positive quantities. Therefore at late phase we get an accelerated expansion with a negative value of deceleration parameter. However, at an early epoch, there is a possibility of deceleration as we have desired for a cosmic transit at certain instant of time for which the functional $F(t)$ may vanish. At an early time, we may neglect the contribution from $3a^2$ compared   to the evolving terms of the functional $F(t)$ which leads to a cosmic time $t=\frac{1-3b}{6a}$ when $F(t)$ vanishes. This further constrains the parameter $b$ in a more tighter range $0<b<\frac{1}{3}$, so that we can only get a positive cosmic time frame to have a decelerated universe. The functional $F(t)$ evolves from a large negative value in early phase to reach to a positive maximum at certain past cosmic time and then decreases to small non zero values at late phase. 

The energy density $\rho$ is obtained as
\begin{equation}
\rho=\frac{2(m^2+4m+1)}{(m+1 )^2}\left(a+\frac{b}{t}\right)^2-3\left(\frac{\alpha}{e^{at}t^b}\right)^2.\label{eq:24}
\end{equation}
At an early cosmic time when $t\longrightarrow 0$, the behaviour of the energy density is mostly decided by the terms involving low power in $t$ i.e. $\rho \sim 3\left[\frac{b^2}{t^2}-\frac{\alpha^2}{t^{2b}}\right]$. More specifically, since $b<<1$, at an early cosmic phase, $\rho \sim \frac{3b^2}{t^2}$, provided the parameter $\alpha$ is not very large. At late cosmic phase, the energy density evolves to become $\rho \simeq \frac{2a^2(m^2+4m+1)}{(m+1 )^2}$. It is now clear that, both at the early and late cosmic times, the energy density is positive which is required for viable cosmological models. However, with the growth of cosmic time, there is a possibility that the energy density may become negative at some point of time because of the dominance of exponential term. In order to avoid such unphysical situation, we restrict the parameters in such a manner that at any time $t$ they should satisfy the condition of $\sqrt{\frac{2(m^2+4m+1)}{3\alpha^2 (m+1 )^2}}\left(a+\frac{b}{t}\right)> e^{-at}t^{-b}$.

The skewness parameters normalised to the functional $\zeta(m)$ can be obtained using equations \eqref{eq:18}-\eqref{eq:20} along with equations \eqref{eq:23} and \eqref{eq:24}as 

\begin{eqnarray}
\delta &=& -\frac{2}{m+1}\left[3a^2+\frac{6ab}{t}+\frac{(3b-1)b}{t^2}\right]\left(\frac{m-1}{3\rho}\right),\label{eq:25}\\
\gamma &=&  \frac{2}{m+1}\left[3a^2+\frac{6ab}{t}+\frac{(3b-1)b}{t^2}\right]\left(\frac{5+m}{6\rho}\right),\label{eq:26}\\
\eta &=& -\frac{2}{m+1}\left[3a^2+\frac{6ab}{t}+\frac{(3b-1)b}{t^2}\right]\left(\frac{5m+1}{6\rho}\right).\label{eq:27}
\end{eqnarray}

The evolutionary behaviour of the skewness parameters are decided by the time varying nature of the factor  $\frac{F(t)}{\rho}$ which becomes negative at early cosmic phase and positive at late times. The skewness parameter along $x-$axis almost does not evolve with time and remains close to zero. This is what we expect earlier since, along $x-$axis, the expansion rate is considered to be the same as that of the mean Hubble rate. May be that is the reason, along this axis, the pressure of the anisotropic fluid is the same as that of the mean pressure. Since, $m$ is close to 1 to handle the little anisotropic nature of the universe, the behaviour of $\gamma$ is found to be just the mirror image of the behaviour of $\eta$. At an early cosmic time, the factor $\frac{F(t)}{\rho}$ behaves like $\frac{(3b-1)(m+1)^2}{m^2+4m+1}$ which assumes a small negative value for $b$ in the range $0<b<\frac{1}{3}$. And at late phase, this factor behaves like $\frac{F(t)}{\rho} \sim \frac{3(m+1)}{m^2+4m+1}$. Consequently $\gamma$ evolves from $\frac{(5+m)(3b-1)(m+1)^2}{6(m^2+4m+1)}$ to become $\frac{(5+m)(m+1)}{2(m^2+4m+1)}$ at late time of cosmic evolution. 

The equation of state parameter for the model is 

\begin{equation}
\omega=-\frac{2(m^2+4m+1)}{3\rho(m+1)^2}\left[-\frac{2b}{t^2}+3\left(a+\frac{b}{t}\right)^2\right]+\frac{3\alpha ^2}{\rho e^{2at}t^{2b}}.\label{eq:28}
\end{equation}

The equation of state parameter evolves dynamically with the expansion of the universe. The dynamics is mostly governed by the behaviour the rest energy density. At an early phase, when the power law behaviour of the scale factor dominates the dynamics, the equation of state parameter behaves like $\omega= -1+\frac{2}{3b}$,  which is a positive quantity for $0<b<\frac{1}{3}$. At a late phase of evolution, it assumes a constant value $\omega \sim -3$. The equation of state parameter evolves from a positive value in the beginning, crosses the phantom divide and at late times enters into phantom region. 

\subsection{Anisotropic behaviour of the model}

The average anisotropy parameter $\mathcal{A}$ is defined as
\begin{equation}\label{eq:15}
\mathcal{A}=\frac{1}{3}\Sigma \left(\frac{\Delta H_i}{H}\right)^2,
\end{equation}
where $\Delta H_i=H_i-H; i=1,2,3.$ $\mathcal{A}$ is a measure of deviation from isotropic expansion. A model is isotropic if $\mathcal{A}=0$, otherwise the model is anisotropic. One should note that, a model isotropizes at late phase of cosmic evolution if the volume scale factor increases to infinitely large value and on the otherhand, the average anisotropic parameter vanishes for large value of cosmic time. In terms of the exponent $m$, the average anisotropic parameter for the present model can be expressed as \cite{BM15} 

\begin{equation}\label{eq:16}
\mathcal{A}=\frac{2}{3}\left(\frac{m-1}{m+1}\right)^2.
\end{equation}
The above equation clearly indicates that, the exponent $m$ takes care of the anisotropic nature of the model. The model becomes isotropic with equal rate of expansion in all spatial directions if $m=1$; otherwise the model will be anisotropic. The average anisotropic parameter in eq.\eqref{eq:16} is time independent implying that the anisotropy in expansion rates is maintained throughout the cosmic evolution. However, one should note that, the universe is observed mostly to be isotropic and any consideration of anisotropy must be taken as a sort of small perturbation in the expansion rates which necessitates that the value of the exponent $m$ should be very close to 1. As has been calculated from some observational bounds in  an earlier work \cite{BM15}, the value of $m$ is around $m=1.0001633$ in a $BV$ model corresponding to an average anisotropy of $4.4439\times 10^{-9}$.

Concerning the anisotropies in the dark energy fluid pressure, one can note that, the skewness parameters evolve with the cosmic dynamics. For a hybrid scale with  power law and exponential factors, the behaviour of the model is dominated by the power law factor at the early phase whereas the exponential factor dominates at the late phase of evolution. In view of this, the evolution of pressure anisotropies follows almost a similar trend to that of the de Sitter model discussed in Ref.\cite{SKT15} at least at late times. The pressure anisotropy increases initially and after some instant of time, it decreases to low values at late times. This is evident from the late time pressure anisotropy in $y-$ axis which is almost a mirror image of that in $z-$axis. The value of $\gamma$ at late times of cosmic evolution becomes $\frac{(5+m)(m+1)}{2(m^2+4m+1)}$ compared to the value at present epoch $\frac{(5+m)}{3\rho_0 (m+1)}\left[3(a^2+b^2)+b(6a-1)\right]$. The value of the pressure anisotropies are decided by the two constant parameters $a$ and $b$ of the hybrid scale factor besides the exponent $m$. The value of $b$ has been constrained in the present work to lie in the range $0<b<\frac{1}{3}$ whereas the value of $a$ can be constrained from the behaviour of the Hubble parameter at different redshifts and the cosmic transit phenomena. However, in the present investigation, we take this as a free parameter.

\section{Statefinder diagnosis}
The viability of dark energy models can be tested through the statefinder diagnostic pair $\{r,s\}$ which provide us an idea about the geometrical nature of the model. The statefinder pair $\{r,s\}$ are defined as 

\begin{eqnarray}
r &=& \frac{\dddot{\mathcal{R}}}{\mathcal{R}H^3},\\ \nonumber
s &=& \frac{r-1}{3(q-\frac{1}{2})}.\label{eq:29}
\end{eqnarray}

For the present dark energy model with anisotropic pressures along different spatial directions and the imposition of a hybrid scale factor to simulate a cosmic transit from decelerating phase to an accelerating one, the statefinder pair can be obtained as

\begin{eqnarray}
r &=& 1-\frac{3b}{(at+b)^2}+\frac{2b}{(at+b)^3}, \\ \nonumber
s &=& \frac{-6b(at+b)+4b}{6b(at+b)-9(at+b)^3}.\label{30}
\end{eqnarray}

The values of the statefinder pair depend on the parameters $a$ and $b$ of the hybrid scale factor chosen. Both $r$ and $s$ evolve with time from large value to small value at late time. At the beginning of cosmic time, the statefinder pair for the present model are $\{1+\frac{2-3b}{b^2}, \frac{2}{3b}\}$ whereas at late time of cosmic evolution, the model behaves like $\Lambda$CDM with the statefinder pair  having values $\{1,0\}$.

\section{Conclusion}

In the present work, we have constructed an anisotropic dark energy cosmological model in the framework of General Relativity at the backdrop of spatially homogeneous and anisotropic Bianchi V metric. The anisotropic behaviour of the model is simulated through the consideration of different scale factors and Hubble expansion rates along different spatial directions. A parameter $m$ is considered to take care of the anisotropic behaviour of the model in the sense that, if $m=1$, we get isotropic model and for $m\neq 1$, anisotropic nature will be retained. The cosmic fluid is also considered to be anisotropic which allow us to assume different pressure of the fluid along different directions. 

The accelerated expansion of the universe is observed to have  occurred at late phase of cosmic dynamics and before this the universe might be decelerating at an early time. In otherwords, the universe might have undergone  a transition from early deceleration to late time acceleration at certain point of time. Such a situation can be simulated through a time varying deceleration parameter which may be positive at early time and evolves to negative values at late times. We consider such a deceleration parameter which can be generated from a scale factor having a hybrid form containing factors of exponential behaviour and power law behaviour. It is worth to mention here that, power law and exponential scale factors are widely used in literature for the investigation of background cosmologies. Also, these two kinds of scale factors lead to a constant deceleration parameter. The power law factor of the hybrid scale factor, we have used in the present work, dominate the early part of cosmic dynamics where as the exponential factor dominate at late times providing a realistic cosmological model.

We followed the general formalism developed for dark energy models with pressure anisotropies in earlier works \cite {BM15,SKT15} to get the evolution of skewness parameters and the equation of state parameter. In the context of the discussed model, we have constrained the parameters of the hybrid scale factors from some observational as well as physical bounds. The interesting feature of the model is that, the skewness parameters dynamically evolve with the cosmic expansion which speaks of a dynamically changing pressure anisotropies along different spatial directions. Along the $x-$axis, the skewness parameter almost remains constant with values close to zero signifying that, the pressure along this direction is equal to that of the mean pressure. As in the previous works, in this investigation, we found that, the pressure anisotropies along $y-$ and $z-$axes behave just as the mirror image of the other. Also, it is observed from the discussed model that, the pressure anisotropies along the $y-$ and $z-$axes continue with the cosmic expansion without being subsided at any point of time.

The equation of state parameter is obtained to  vary with cosmic time implying an evolving relationship between the mean pressure of the cosmic fluid with the energy density through out its evolution. It evolves from a positive quantity at the beginning to enter into a phantom region at late times. We have also calculated the statefinder parameters to test the dark energy model. The statefinder pair also come out to be time varying and they decrease with the cosmic evolution to overlap with the $\Lambda$CDM model at late times. The constructed model being more realistic to simulate a cosmic transit favours a phantom phase at late times. The use of a hybrid scale factor significantly changes the behaviour of the cosmic fluid. 

\section{Acknowledgement}
BM acknowledges SERB-DST, New Delhi, India for financial support to carry out the Research project [No.-SR/S4/MS:815/13]. BM and SKT acknowledge the hospitality of IUCAA, Pune(India) during an academic visit where a part of this work is done.

\end{document}